\documentclass[twocolumn,showpacs,preprintnumbers,amsmath,amssymb]{revtex4}
\usepackage{graphics,epsfig,subfigure}
\usepackage{ulem}
\usepackage{color}

\begin{document}
\newcommand{\PR}[1]{\ensuremath{\left[#1\right]}} 
\newcommand{\PC}[1]{\ensuremath{\left(#1\right)}} 
\newcommand{\PX}[1]{\ensuremath{\left\lbrace#1\right\rbrace}} 
\newcommand{\BR}[1]{\ensuremath{\left\langle#1\right\vert}} 
\newcommand{\KT}[1]{\ensuremath{\left\vert#1\right\rangle}} 
\newcommand{\MD}[1]{\ensuremath{\left\vert#1\right\vert}} 

\title{Weak deflection angle by electrically and magnetically charged black holes from nonlinear electrodynamics}
\author{Qi-Ming Fu$^{1,}$\footnote{fuqiming@snut.edu.cn},
        Li Zhao$^{2,}$\footnote{lizhao@lzu.edu.cn},
        and Yu-Xiao Liu$^{2,3,}$\footnote{liuyx@lzu.edu.cn, corresponding author}}

\affiliation{$^{1}$Institute of Physics, Shannxi University of Technology, Hanzhong 723000, China \\
             $^{2}$Institute of Theoretical Physics $\&$ Research Center of Gravitation, Lanzhou University, Lanzhou 730000, China \\
             $^{3}$Lanzhou Center for Theoretical Physics, Key Laboratory of Theoretical Physics of Gansu Province, School of Physical Science and Technology, Lanzhou University, Lanzhou 730000, China}

\begin{abstract}
  Nonlinear electrodynamic (NLED) theories are well-motivated for their extensions to classical electrodynamics in the strong field regime, and have been extensively investigated in seeking for regular black hole solutions. In this paper, we focus on two spherically symmetric and static black hole solutions based on two types of NLED models: the Euler-Heisenberg NLED model and the Bronnikov NLED model, and calculate the weak deflection angle of light by these two black holes with the help of the Gauss-Bonnet theorem. We investigate the effects of the one-loop corrections to quantum electrodynamics on the deflection angle and analyse the behavior of the deflection angle by a regular magnetically charged black hole. It is found that the weak deflection angle of the electrically charged Einstein-Euler-Heisenberg black hole increases with the one-loop corrections and the regular magnetically charged black hole based on the Bronnikov NLED model has a smaller deflection angle than the singular one. Besides, we also calculate the deflection angle of light by the geodesic method for verification. In addition, we discuss the effects of a cold non-magnetized plasma on the deflection angle and find that the deflection angle increases with the plasma parameter.
\end{abstract}

\pacs{95.30.Sf, 98.62.Sb, 97.60.Lf}




\maketitle

\section{Introduction}

To solve the divergence of the self energy of a point-like charge, Born and Infeld generalized Maxwell's theory and proposed the Born-Infeld electrodynamics \cite{Born1934}. However, this theory did not attract much attention until its reemergence at the low energy scale of some string theories. Afterwards, Heisenberg and Euler introduced a new extension to the standard electromagnetic theory (known as Euler-Heisenberg (EH) electrodynamics) \cite{Heisenberg1936}, which takes into account the one-loop corrections to quantum electrodynamics (QED) and can explain the vacuum polarization in QED. As extensions to Born-Infeld and EH electrodynamics, nonlinear electrodynamic (NLED) models have been studied in different aspects since then. For instance, NLED models can be used to explain the inflation of the universe in the early times \cite{Salcedo2000,Camara2004}. Some types of NLED models can depict the accelerated expansion of the universe instead of dark energy and remove the Big Bang singularity \cite{Elizalde2003,Novello2004,Vollick2008,Kruglov2015}.

In addition, in recent years, NLED models attract much more attention for their ability in seeking regular black hole solutions. The first regular black hole model was proposed by Bardeen \cite{Bardeen1968}. However, this regular black hole was obtained without a specified source associated to its line element. Remarkably, in 1998, Ay$\acute{\text{o}}$n-Beato and Garc$\acute{\text{{\i}}}$a found that NLED model minimally coupled to general relativity (GR) can be a possible source generating such a regular black hole solution \cite{Ayon-Beato1998}. In Ref.~\cite{Bronnikov2001}, Bronnikov found a class of magnetically charged regular black holes in the framework of GR coupled with a specific NLED model (known as Bronnikov NLED model). Subsequently, Hayward proposed a concrete model which can describe both the collapse and evaporation of black holes \cite{Hayward2006}. One can see Refs.~\cite{Dymnikova1992,Ayon-Beato2000,Elizalde2002,Nicolini2006,Ansoldi2007,Hossenfelder2010,Johannsen2013,Dymnikova2015,Rodrigues2016,Fan2016,Chinaglia2017,Nojiri2017,Yu2020,Pedro2020} for more regular black holes based on NLED models. In this paper, we mainly focus on two black hole models based on two particular above-mentioned NLED models, i.e., EH NLED model and Bronnikov NLED model, and investigate the weak deflection angle of light by these two black hole models.

On the other hand, it is well known that light rays will be bent when traveling through a massive object, known as the gravitational lensing effect, which is one of the key predictions of GR. At present, the gravitational lensing is one of the most powerful tools in astronomy and cosmology, such as, measuring the mass of galaxies and clusters \cite{Hoekstra2013,Brouwer2018,Bellagamba2019}, detecting dark energy and dark matter \cite{Vanderveld2012,He2017,Cao2012,Huterer2018,Jung2019,Andrade2019,Turimov2019}. Since the first measurement of the gravitational bending of light by the sun, the gravitational lensing effects have been extensively investigated for black holes, wormholes, cosmic strings and other objects by the lens equation \cite{Keeton1998,Bhadra2003,Perlick2004,Whisker2005,Chen2009,Nandi2006,Eiroa2002,Mao1991,Bozza2002,Hoekstra2004,Virbhadra2002,Virbhadra2000,Gallo2011,Sharif2015,Gibbons1993}. In 2008, Gibbons and Werner introduced an alternative method to calculate the weak deflection angle of light in static asymptotically flat spacetimes by using the Gauss-Bonnet theorem and the optical geometry of the spacetime, where the light source and receiver are located at infinity \cite{Gibbons2008}.
Later, this method was extended to stationary spacetimes by Werner \cite{Werner2012}. In Ref.~\cite{Ishihara2016}, the authors investigated the weak deflection of light for the light source and receiver located at a finite distance. The weak deflection for the massive particles by this method was investigated in Refs.~\cite{Crisnejo2018,Jusufi2018,Zonghai2020}. Besides, the weak deflection of light by a black hole immersed in a plasma medium was discussed in Ref.~\cite{Crisnejo2018}. One can see Refs.~\cite{Jusufi2016,Jusufi2017a,Jusufi2017b,Ono2017,Sakalli2017,Jusufi2018a,Jusufi2018b,Jusufi2018c,Arakida2018,Ono2018,Gyulchev2019,Javed2019,Sakalli2019,Crisnejo2019}
for more recent works.

Although the black holes based on Einstein-Euler-Heisenberg (EEH) theory have been extensively studied in the literatures \cite{Yajima2001,Ruffini2013,Guerrero2020,Allahyari2020,Magos2020}, the weak deflection of light by these black holes have not been investigated yet. As a powerful tool to study the characteristics of black holes, it is interesting to investigate the weak deflection angle by the electrically charged EEH black hole and know what the effects are of the one-loop corrections to QED on the deflection angle. Besides, although there are many investigations on the NLED-based regular black holes, the weak deflection angle of light by such regular black holes are rarely investigated. In this paper, we take the Bronnikov NLED black hole with magnetic charge as an example and investigate the characteristics of this regular black hole by calculating its deflection angle. What's more, most astrophysical objects including black holes are surrounded by a plasma medium. Thus, it is interesting to investigate the effects of the plasma medium on the deflection angle of light by these black holes.

This paper is organized as follows. In Sec.~\ref{Euler}, we first give a brief review of the EEH black hole and then calculate the weak deflection angle of light by this black hole via two different methods, i.e., the method by using the Gauss-bonnet theorem and the traditional geodesic method. Then, the effects of the plasma on the weak deflection angle are studied. In Sec.~\ref{Bronnikov}, we perform the same procedures for the Bronnikov NLED black hole and analyse the characteristics of the weak deflection angle of light by this regular magnetically charged black hole. Section \ref{con} comes with the conclusion.

\section{Weak deflection angle of light by the Einstein-Euler-Heisenberg black holes}~\label{Euler}

In this section, we first give a brief review of the Einstein-Euler-Heisenberg theory and present the spherically symmetric and static solution to this theory. Then, we will use these results to calculate the weak deflection angle of light for this black hole by using the Gauss-Bonnet theorem. Besides, the weak deflection angle of light is also calculated with the null geodesic method as a verification to the former results. Finally, we will investigate the deflection angle of light for this black hole immersed in a cold non-magnetized plasma medium.

\subsection{Einstein-Euler-Heisenberg theory}

The action for the Einstein-Euler-Heisenberg theory is given by \cite{Allahyari2020,Magos2020}
\begin{eqnarray}
S=\frac{1}{4\pi}\int d^4x\sqrt{-g}\left[\frac{1}{4}R-\mathcal{L}(F,G)\right],
\end{eqnarray}
where $\mathcal{L}(F,G)$ is the functional of the electromagnetic invariants, $F=\frac{1}{4}F_{\mu\nu}F^{\mu\nu}$ and $G=\frac{1}{4}F^{\mu\nu}F^{*}_{\mu\nu}$ with $F_{\mu\nu}$ the electromagnetic field strength and $F^{*}_{\mu\nu}=\frac{1}{2}\epsilon_{\mu\nu\sigma\rho}F^{\sigma\rho}$ its dual. The Levi-Civita tensor satisfies $\epsilon_{\mu\nu\sigma\rho}\epsilon^{\mu\nu\sigma\rho}=-4!$. As one-loop corrections to quantum electrodynamics (QED), the Euler-Heisenberg Lagrangian is
\begin{eqnarray}
\mathcal{L}(F,G)=-F+\frac{a}{2}F^2+\frac{7a}{8}G^2,
\end{eqnarray}
where $a$ is the Euler-Heisenberg parameter. For $a=0$, the standard Maxwell electrodynamics is recovered. There are two frameworks in nonlinear electrodynamics. One is the $F$ framework constructed by the electromagnetic field tensor $F_{\mu\nu}$ and the other is the $P$ framework constructed by the tensor $P_{\mu\nu}$, defined by
\begin{eqnarray}
P_{\mu\nu}=-(\mathcal{L}_F F_{\mu\nu}+F^{*}_{\mu\nu}\mathcal{L}_G),
\end{eqnarray}
where $\mathcal{L}_X=\frac{\partial\mathcal{L}}{\partial X}$. Then, the $P_{\mu\nu}$ in the Euler-Heisenberg theory can be calculated as
\begin{eqnarray}
P_{\mu\nu}=(1-aF)F_{\mu\nu}-\frac{7a}{4}F^{*}_{\mu\nu}G.
\end{eqnarray}

In the $P$ framework, one can define two independent invariants $P$ and $O$,
\begin{eqnarray}
P=-\frac{1}{4}P_{\mu\nu}P^{\mu\nu}, \quad\quad O=-\frac{1}{4}P^{\mu\nu}P^{*}_{\mu\nu},
\end{eqnarray}
where $P^{*}_{\mu\nu}=\frac{1}{2}\epsilon_{\mu\nu\sigma\rho}P^{\sigma\rho}$.

The equations of motion can be derived as
\begin{eqnarray}
R_{\mu\nu}-\frac{1}{2}g_{\mu\nu}R&=&8\pi T_{\mu\nu}, \\
\nabla_{\mu}P^{\mu\nu}&=&0,
\end{eqnarray}
where the energy momentum tensor in the $P$ framework is given by
\begin{eqnarray}
T_{\mu\nu}\!\!=\!\!\frac{1}{4\pi}\left[(1\!-\!aP)P_{\mu}^{\sigma}P_{\nu\sigma}\!+\!g_{\mu\nu}\left(P\!-\!\frac{3}{2}aP^2\!-\!\frac{7a}{8}O^2\right)\right]. \nonumber\\
\end{eqnarray}

\subsection{Spherically symmetric solution in the Einstein-Euler-Heisenberg theory}

The line element for a spherically symmetric and static black hole can be assumed as
\begin{eqnarray}~\label{metric}
ds^2=g_{\mu\nu}dx^{\mu}dx^{\nu}=-f(r)dt^2+f(r)^{-1}dr^2+r^2d\Omega^2,
\end{eqnarray}
where $\mu$ and $\nu$ run from $0$ to $3$, and $d\Omega^2=d\theta^2+\sin^2\theta d\phi^2$.
According to the symmetry of the spacetime and restricting to the electric charge $Q$, the $P_{\mu\nu}$ can be calculated as
\begin{eqnarray}
P_{\mu\nu}=\frac{Q}{r^2}\delta^0_{[\mu}\delta^1_{\nu]},
\end{eqnarray}
and the independent electromagnetic invariants are
\begin{eqnarray}
P=\frac{Q^2}{2r^4}, \quad\quad O=0.
\end{eqnarray}
Then the function in the metric can be solved as \cite{Allahyari2020,Magos2020}
\begin{eqnarray}
f(r)=1-\frac{2M}{r}+\frac{Q^2}{r^2}-\frac{aQ^4}{20r^6},
\end{eqnarray}
where $M$ is the mass of the black hole.

\subsection{Calculation of deflection angle with the Gauss-Bonnet theorem}

The null geodesics satisfies $ds^2=0$, which can be rearranged as
\begin{eqnarray}
dt^2=\gamma_{ij}dx^i dx^j=\frac{1}{f^2}dr^2+\frac{r^2}{f}d\Omega^2,~\label{opmetric}
\end{eqnarray}
where $i$ and $j$ run from $1$ to $3$, and $\gamma_{ij}$ is the so-called optical metric. After a coordinate transformation $dr^*=\frac{1}{f}dr$, the above expression can be rewritten as
\begin{eqnarray}
dt^2=dr^{*2}+\tilde{f}^2(r^*)d\phi^2,
\end{eqnarray}
where $\tilde{f}(r^*)\equiv\sqrt{\frac{r^2}{f}}$ and $\theta=\frac{\pi}{2}$.
The Gaussian curvature of the optical spacetime can be calculated as
\begin{eqnarray}
\mathcal{K}&=&\frac{R_{r\phi r\phi}}{\gamma}=\frac{1}{\sqrt{\gamma}}\left[\frac{\partial}{\partial \phi}\left(\frac{\sqrt{\gamma}}{\gamma_{rr}}\Gamma^{\phi}_{rr}\right)-\frac{\partial}{\partial r}\left(\frac{\sqrt{\gamma}}{\gamma_{rr}}\Gamma^{\phi}_{r\phi}\right)\right] \nonumber\\
&=&-\frac{2M}{r^3}\left(1-\frac{3 M}{2r}\right)+\frac{Q^2}{r^4}\left(3-\frac{6 M}{r}\right) \nonumber\\
&+&\frac{Q^4}{r^6}\left(2-\frac{21 a}{20 r^2}+\frac{19 a M}{10 r^3}\right)-\frac{9 a Q^6}{10 r^{10}}+\frac{3 a^2 Q^8}{100 r^{14}},~\label{GC}
\end{eqnarray}
where $\gamma\equiv\det (\gamma_{ij})$.

Let the domain $D$ be a compact oriented nonsingular two-dimensional Riemannian surface with Euler characteristic $\chi(D)$ and Gaussian curvature $\mathcal{K}$, and bounded by a piecewise smooth curve with geodesic curvature $\kappa$. Then the Gauss-Bonnet theorem gives the relation between the deflection angle of light and the Gaussian curvature via
\begin{eqnarray}
\int\int_D \mathcal{K}dS+\oint_{\partial D}\kappa dt+\sum_{i=1}\beta_i=2\pi \chi(D),~\label{GB}
\end{eqnarray}
where $dS$ is the surface element, $\kappa$ standards for the geodesic curvature of the boundary defined as $\kappa=|\nabla_{\dot{C}}\dot{C}|$, and $\beta_i$ denotes the $i^{\text{th}}$ exterior angles. For a specific $\tilde{D}$ bounded by a geodesic $C_1$ from the source $S$ to the observer $O$ and a circular curve $C_R$ intersecting $C_1$ in $S$ and $O$ at right angles, Eq.~(\ref{GB}) reduces to
\begin{eqnarray}
\int\int_{\tilde{D}}\mathcal{K}dS+\int_{C_R}\kappa(C_R)dt=\pi,~\label{GB2}
\end{eqnarray}
where we have used $\kappa(C_1)=0$ and the Euler characteristic $\chi(\tilde{D})=1$. For the circular curve $C_R := r(\phi)=R=\text{const}$, the non-zero part of the geodesic curvature can be calculated as
\begin{eqnarray}
\kappa(C_R)=\left(\nabla_{\dot{C}_R}\dot{C}_R\right)^r=\dot{C}^{\phi}_R(\partial_{\phi}\dot{C}^r_R)+\Gamma^r_{\phi\phi}(\dot{C}^{\phi}_R)^2,
\end{eqnarray}
where $\dot{C}_R$ denotes the tangent vector of the circular curve $C_R$ and $\Gamma^r_{\phi\phi}$ is the Christoffel symbol related to the optical metric (\ref{opmetric}). In the last equation it is obvious that the first term vanishes, and $\Gamma^r_{\phi\phi}=-\tilde{f}(r^*)\tilde{f}'(r^*)$, $(\dot{C}^{\phi}_R)^2=\frac{1}{\tilde{f}^2(r^*)}$ in the second term. In the limit $R\rightarrow \infty$, one can obtain
\begin{eqnarray}
&&\lim_{R\rightarrow \infty}\left[\kappa(C_R)dt\right] \nonumber\\
&=&\lim_{R\rightarrow \infty}[-\tilde{f}'(r^*)]d\phi \nonumber\\
&=&\lim_{R\rightarrow \infty}\left(\frac{10 R^4 \left(R (R-3 M)+2 Q^2\right)-2 a Q^4}{R^3 \sqrt{100 R^4 \left(R (R-2 M)+Q^2\right)-5 a Q^4}}\right)d\phi \nonumber\\
&=&d\phi.~\label{geoR}
\end{eqnarray}

Inserting Eq.~(\ref{geoR}) into Eq.~(\ref{GB2}), one has
\begin{eqnarray}
\int\int_{\tilde{D}_{R\rightarrow \infty}}\mathcal{K}dS+\int_0^{\pi+\alpha}d\phi=\pi.
\end{eqnarray}
Then the weak deflection angle of light can be calculated as
\begin{eqnarray}
\alpha&=&-\int\int_{\tilde{D}}\mathcal{K}dS=-\int^{\pi}_0\int^{\infty}_{\frac{b}{\sin\phi}}\mathcal{K}dS \nonumber\\
&\simeq&\frac{4 M}{b}-\frac{3 \pi  Q^2}{4 b^2}+\frac{7 \pi  a Q^4}{128 b^6}+\mathcal{O}(M^2,a^2,Q^4),~\label{deflang}
\end{eqnarray}
where we have used the zero-order particle trajectory $r=b/\sin\phi$, $0\leq\phi\leq\pi$ at the weak deflection limit.
It is obvious that the first two terms are the deflection angle of light by an electrically charged black hole based on the standard electrodynamics \cite{Jusufi2016}.
The third term
comes from the influences of the one-loop corrections to QED on the spacetime of the black hole. It is obvious that the deflection angle increases with the one-loop corrections while their effects are suppressed by the impact parameter.

\subsection{Calculation of deflection angle by the geodesic method}

The Lagrangian of the null geodesics of the Einstein-Euler-Heisenberg black hole is given by
\begin{eqnarray}
2\mathcal{L}_{*}=-f(r)\dot{t}^2+f(r)^{-1}\dot{r}^2+r^2\big(\dot{\theta}^2+\sin^2\theta\dot{\phi}^2\big),~\label{lagrangian}
\end{eqnarray}
where $\dot{x}=\frac{dx}{d\tau}$, and $\tau$ is the affine parameter along the geodesics. Since the Lagrangian is independent on $t$ and $\phi$, one can obtain two conserved constants:
\begin{eqnarray}
p_{t}&=&\frac{\partial\mathcal{L}_*}{\partial \dot{t}}=-f(r)\dot{t}=-E, \\
p_{\phi}&=&\frac{\partial\mathcal{L}_*}{\partial\dot{\phi}}=r^2\dot{\phi}\sin^2\theta=L.
\end{eqnarray}
Then the null geodesic equation at the equatorial plane can be obtained as
\begin{eqnarray}
\left(\frac{d\phi}{dr}\right)^2=\left(\frac{r^4}{b^2}-r^2f(r)\right)^{-1},~\label{geoeq}
\end{eqnarray}
where the impact parameter is defined as $b=r_0/\sqrt{f(r_0)}$ with $r_0$ the radius of the circular orbit.

The weak bending angle of the light coming from infinity and deflected by a black hole before arriving at infinity is given by
\begin{eqnarray}
\alpha(r_0)=\Delta\phi(r_0)-\pi,
\end{eqnarray}
where $\Delta\phi(r_0)$ can be solved from Eq.~(\ref{geoeq}) as
\begin{eqnarray}
\Delta\phi(r_0)=2\int_{r_0}^{\infty}\left(\frac{r^4}{b^2}-r^2f(r)\right)^{-\frac{1}{2}}dr.~\label{dphi}
\end{eqnarray}

It is convenient to define the dimensionless line element as
\begin{eqnarray}
dS^2&=&(2M)^{-2}ds^2=-f(x)dT^2+f(x)^{-1}dx^2 \nonumber\\
    &+&x^2(d\theta^2+\sin^2\theta d\phi^2),
\end{eqnarray}
where we have defined
\begin{eqnarray}
x=\frac{r}{2M}, \quad T=\frac{t}{2M}, \quad q=\frac{Q}{2M}, \quad \hat{\alpha}=\frac{a}{(2M)^2},
\end{eqnarray}
and the function $f(r)$ in the metric (\ref{metric}) can be reexpressed as
\begin{eqnarray}
f(x)=1-\frac{1}{x}+\frac{q^2}{x^2}-\frac{\hat{\alpha}  q^4}{20 x^6}.
\end{eqnarray}

Then Eq.~(\ref{dphi}) can be rewritten as
\begin{eqnarray}
\Delta\phi(x_0)\!\!&=&\!\!2 \int_{x_0}^{\infty }\!\!\sqrt{20}x^2 x_0^4 \Big(\hat{\alpha}  q^4 \left(x_0^8-x^8\right) \nonumber\\
 \!\!&+&\!\!20 q^2 x^4 x_0^4 \left(x^4-x_0^4\right) \nonumber\\
 \!\!&+&\!\!20 x^5 x_0^5 \left(x^3 (x_0\!-\!1)\!-\!x x_0^3\!+\!x_0^3\right)\Big)^{-\frac{1}{2}}\!\! dx,
\end{eqnarray}
and the impact parameter can be expressed as
\begin{eqnarray}
\frac{b}{2M}=\frac{x_0}{\sqrt{f(x_0)}}.~\label{im}
\end{eqnarray}

After defining a new variable $z=\frac{x_0}{x}$, the above integral can be rewritten as
\begin{eqnarray}
\Delta\phi(x_0)&=&2 \int_0^1 \sqrt{20}x_0^3\Big(\hat{\alpha}  q^4 \left(z^8-1\right)-20 q^2 x_0^4 \left(z^4-1\right) \nonumber\\
 &-&20 x_0^5 \left(x_0 \left(z^2-1\right)-z^3+1\right)\Big)^{-\frac{1}{2}} dz.
\end{eqnarray}
Considering the weak gravitational lensing limit $x_0\gg 1$ and expanding the above integrand about $\frac{1}{x_0}$, the above integral can be integrated out term by term as follows:
\begin{widetext}
\begin{eqnarray}
\alpha(x_0)&=&\Delta\phi(x_0)-\pi=\frac{2}{x_0}+\left(\frac{\pi}{4} \left(\frac{15}{4}-3 q^2\right)-1\right)\frac{1}{x_0^2}+\left(\frac{3}{16} \pi  \left(4 q^2-5\right)-7 q^2+\frac{61}{12}\right)\frac{1}{x_0^3}+\bigg(\frac{5}{8} \left(20 q^2-13\right) \nonumber\\
&+&\frac{3 \pi  \left(304 q^4-2200 q^2+1155\right)}{1024}\bigg)\frac{1}{x_0^4}+\left(\frac{1}{32} (632-105 \pi ) q^4+\frac{7}{64} (135 \pi -536) q^2+\frac{7783}{320}-\frac{3465 \pi }{512}\right)\frac{1}{x_0^5} \nonumber\\
&+&\bigg(\frac{7}{128} \pi  \alpha  q^4-\frac{1}{384} \left(28560 q^4-59832 q^2+21397\right)-\frac{105 \pi }{16384}\left(192 q^6-4816 q^4+8676 q^2-2959\right)\bigg)\frac{1}{x_0^6} \nonumber\\
&+&\mathcal{O}\left(\frac{1}{x_0^7}\right).~\label{alph}
\end{eqnarray}
\end{widetext}

To obtain the deflection angle in terms of the impact parameter $b$, one needs the relation between $b$ and $x_0$ which can be solved from Eq.~(\ref{im}) in the weak deflection limit as
\begin{eqnarray}
\frac{1}{x_0}\!\!&=&\!\!\frac{2M}{b}+\frac{1}{2}\left(\frac{2M}{b}\right)^2-\frac{1}{8}(4q^2-5)\left(\frac{2M}{b}\right)^3-\bigg(\frac{3q^2}{2} \nonumber\\
\!\!&-&\!\!1\bigg)\left(\frac{2M}{b}\right)^4+\frac{7}{128}\Big(16q^4-72q^2+33\Big)\left(\frac{2M}{b}\right)^5 \nonumber\\
\!\!&+&\!\!\left(5q^4\!-\!10q^2\!+\!\frac{7}{2}\right)\left(\frac{2M}{b}\right)^6\!+\!\mathcal{O}\left(\left(\frac{2M}{b}\right)^7\right).
~\label{x0b}
\end{eqnarray}

Inserting Eq.~(\ref{x0b}) into Eq.~(\ref{alph}), the weak deflection angle is found to be
\begin{eqnarray}
\hat{\alpha}\simeq\frac{4M}{b}-\frac{3\pi Q}{4b^2}+\frac{7\pi a Q^4}{128b^6}+\mathcal{O}(M^2,a^2,Q^4).
\end{eqnarray}
It is obvious that the above result is in agreement with the result calculated by using the Gauss-Bonnet theorem. However, it should be noted that this agreement only holds for the first-order terms and breaks down for the higher-order corrections.

\subsection{Weak deflection angle in the presence of plasma}

In this subsection, we investigate the effects of a cold non-magnetized plasma on the deflection angle for the Einstein-Euler-Heisenberg black hole. The refractive index for this black hole is given by \cite{Perlick2015},
\begin{eqnarray}
n(r)=\sqrt{1-\frac{\omega_e^2}{\omega_{\infty}^2} f(r)},
\end{eqnarray}
where $\omega_e$ and $\omega_{\infty}$ denote the electron plasma frequency and the photon frequency measured by a static observer at infinity, respectively. The corresponding optical line element can be defined as
\begin{eqnarray}
d\sigma^2&=&\gamma_{ij}dx^i dx^j=-\frac{n^2}{g_{00}}g_{ij}dx^i dx^j \nonumber\\
&=&n^2\left(\frac{1}{f^2}dr^2+\frac{r^2}{f}d\phi^2\right),~\label{oppmetric}
\end{eqnarray}
which is conformally related to the induced metric on the spatial section with $\theta=\frac{\pi}{2}$. Then the Gaussian curvature can be calculated as
\begin{eqnarray}
\tilde{\mathcal{K}}\!\!&=&\!\!{40 \Xi^{-3} r^4 \left(3 a Q^4+20 r^4 \left(M r-Q^2\right)\right)^2} \nonumber\\
\!\!&+&\!\!\Xi^{-1}\left(\frac{3 a Q^4}{20 r^8}+r^{-4} \left(M r-Q^2\right)\right) \Big(a Q^4 \nonumber\\
\!\!&-&\!\!20 r^4 \left(r (r-2 M)+Q^2\right)\Big)-\Xi^{-2}\bigg(\frac{9 a^2 Q^8}{r^2} \nonumber\\
\!\!&+&\!\!20 a Q^4 r^2 \left(r (18 r-19 M)+2 Q^2\right) \nonumber\\
\!\!&+&\!\!400 r^6\! \left(-Q^2 r (M\!+\!2 r)\!+\!M r^2 (M\!+\!r)\!+\!Q^4\right)\!\!\bigg),
\end{eqnarray}
where $\Xi=\left(a \delta  Q^4+20 r^4 \left(r^2-\delta  \left(r (r-2 M)+Q^2\right)\right)\right)$ and the plasma parameter is defined by $\delta\equiv \frac{\omega_e^2}{\omega_{\infty}^2}$. For a photon can propagate in the plasma, one should require $\omega_{\infty}\geq\omega_e$, thus $0\leq\delta\leq1$. For more details about the plasma, one can refer to Ref.~\cite{Bisnovatyi-Kogan2010}.
Besides, it follows from Eq.~(\ref{oppmetric}) that
\begin{eqnarray}
\frac{d\sigma}{d\phi}\bigg|_{\gamma_R}=n\sqrt{\frac{r^2}{f}},
\end{eqnarray}
which results in
\begin{eqnarray}
\lim_{R\rightarrow \infty}\tilde{\kappa}(C_R) \frac{d\sigma}{d\phi}\bigg|_{\gamma_R}\approx 1.
\end{eqnarray}

By taking the zero-order particle trajectory $r=\frac{b}{\sin\phi}$ and for the limit $R\rightarrow \infty$, the Gauss-Bonnet theorem can be written as
\begin{eqnarray}
\int^{\pi+\alpha}_0 d\phi=\pi-\int^{\pi}_0\int^{\infty}_{\frac{b}{\sin\phi}}\tilde{\mathcal{K}}dS.
\end{eqnarray}
Then the deflection angle can be calculated as
\begin{eqnarray}
\alpha&=&-\int^{\pi}_0\int^{\infty}_{\frac{b}{\sin\phi}}\tilde{\mathcal{K}}dS \nonumber\\
      &\simeq&\frac{2 M}{b}\left(1+\frac{1}{1-\delta }\right)-\frac{ \pi Q^2}{4 b^2}\left(1+\frac{2}{1-\delta}\right) \nonumber\\
      &+&\frac{a\pi Q^4}{128 b^6}\left(1+\frac{6}{1-\delta}\right)+\mathcal{O}(M^2,a^2,Q^4).~\label{deflangp}
\end{eqnarray}
It can be easily shown that Eq.~(\ref{deflangp}) reduces to Eq.~(\ref{deflang}) when $\delta\rightarrow 0$, and the deflection angle increases with the plasma parameter $\delta$, which suggests that the lower the photon frequency measured by a static observer at infinity is, the larger the deflection angle of it will be for a fixed electron plasma frequency.

\section{Weak deflection angle of light by Einstein-Bronnikov black holes}~\label{Bronnikov}

In this section we will perform the same procedures of the previous section in the case of Einstein-Bronnikov theory, which is a particular NLED theory only consists of the relativistic invariant $F$, and wherein one can obtain regular black holes.

\subsection{The Einstein-Bronnikov theory}

The action for the Einstein-Bronnikov theory is given by \cite{Bronnikov2001}
\begin{eqnarray}
S=\frac{1}{16\pi}\int d^4x\sqrt{-g}\left[R-\mathcal{L}(F)\right],
\end{eqnarray}
where
\begin{eqnarray}
\mathcal{L}(F)=F\cosh^{-2}\left[\hat{a}\left(F/2\right)^{1/4}\right],
\end{eqnarray}
and the parameter $\hat{a}$ is related to the black hole mass $M$ and magnetic charge $Q_m$ via $\hat{a}=Q_m^{3/2}/(2M)$.
The standard Einstein-Maxwell Lagrangian can be recovered with $\hat{a}\rightarrow 0$.

The equations of motion can be derived as
\begin{eqnarray}
R_{\mu\nu}\!-\!\frac{1}{2}g_{\mu\nu}R\!\!&=&\!\!8\pi T_{\mu\nu}\!\!=\!\!8\pi\!\left(2\mathcal{L}_F F_{\rho\mu}F_{\nu}^{\rho}\!-\!\frac{1}{2}g_{\mu\nu}\mathcal{L}\right)\!, \\
\nabla_{\mu}(\mathcal{L}_F F^{\mu\nu})&=&0.
\end{eqnarray}

Considering the spherically symmetric and static spacetime and restricting to the magnetic charge $Q_m$, the relevant function in the metric analogous to Eq.~(\ref{metric}) can be obtained as \cite{Bronnikov2001}
\begin{eqnarray}~\label{metric2}
g(r)=1-\frac{2M}{r}\left(1-\tanh\left(\frac{Q_m^2}{2Mr}\right)\right),
\end{eqnarray}
and the gauge field is given by $A_{\mu}=-Q_m\cos\theta \delta^{\phi}_{\mu}$.
It can be straightforwardly shown that the metric function (\ref{metric2}) reduces to the Schwarzschild black hole solution with $Q_m\rightarrow 0$ and is regular as $r\rightarrow 0$, which suggests a regular black hole.

\subsection{Calculation of deflection angle by the Gauss-Bonnet theorem}

The null geodesics satisfying $ds^2=0$ can be rearranged as
\begin{eqnarray}
dt^2=\gamma_{ij}dx^i dx^j=\frac{1}{g^2}dr^2+\frac{r^2}{g}d\Omega^2.
\end{eqnarray}
 After a coordinate transformation $dr^*=\frac{1}{g}dr$, the above line element can be rewritten as
\begin{eqnarray}
dt^2=dr^{*2}+\tilde{g}^2(r^*)d\phi^2,
\end{eqnarray}
where $\tilde{g}(r^*)=\sqrt{\frac{r^2}{g}}$ and $\theta=\frac{\pi}{2}$.
The Gaussian curvature of this optical spacetime can be calculated as
\begin{widetext}
\begin{eqnarray}
\mathcal{K}&=&-\frac{2M}{r^3}\left(1-\tanh\left(\frac{Q_m}{2Mr}\right)\right)+\frac{1}{r^4}\bigg[3M^2\left(1-\tanh \left(\frac{Q_m^2}{2 M r}\right)\right)^2+2Q_m^2 \text{sech}^2\left(\frac{Q_m^2}{2Mr}\right)\bigg] \nonumber\\
           &-&\frac{Q_m^2}{2Mr^5}\text{sech}^2\left(\frac{Q_m^2}{2Mr}\right)\bigg[6M^2+(Q_m^2-6M^2)\tanh\left(\frac{Q_m^2}{2Mr}\right)\bigg] \nonumber\\
           &-&\frac{Q_m^4}{4r^6}\text{sech}^2\left(\frac{Q_m^2}{2Mr}\right)\left(1-\tanh\left(\frac{Q_m}{2Mr}\right)\right)\left(1-3\tanh\left(\frac{Q_m}{2Mr}\right)\right).~\label{GC2}
\end{eqnarray}
\end{widetext}

Following the same procedures as the previous section, the weak deflection angle of light by this black hole can be obtained as
\begin{eqnarray}
\alpha&=&-\int\int_{\tilde{D}}\mathcal{K}dS=-\int^{\pi}_0\int^{\infty}_{\frac{1}{u(\phi)}}\mathcal{K}dS \nonumber\\
&\simeq&\frac{4 M}{b}-\frac{3 \pi  Q_m^2}{4 b^2}-\frac{16MQ_m^2}{b^3}+\mathcal{O}(M^2,Q_m^3),~\label{deflang2}
\end{eqnarray}
where $u(\phi)$ is given in Eq.~(\ref{uphi}). It is obvious that the first two terms are the same with the weak deflection angle of light by the Reissner-Nordstr$\ddot{\text{o}}$m black hole \cite{Jusufi2016} except the electric charge is replaced by the magnetic charge, and the minus sign in front of the third term indicates that the weak deflection angle of this regular magnetically charged black hole is smaller than the singular one.

\subsection{Calculation of deflection angle by the geodesic method}

The Lagrangian of the null geodesics of the Einstein-Bronnikov black hole is given by
\begin{eqnarray}
2\mathcal{L}_{*}=-g(r)\dot{t}^2+g(r)^{-1}\dot{r}^2+r^2\big(\dot{\theta}^2+\sin^2\theta\dot{\phi}^2\big),~\label{lagrangian}
\end{eqnarray}
where $\dot{x}=\frac{dx}{d\tau}$, and $\tau$ is the affine parameter along the geodesic.
Then the null geodesic equation at the equatorial plane can be obtained as
\begin{eqnarray}
\left(\frac{d\phi}{dr}\right)^2=\left(\frac{r^4}{b^2}-r^2g(r)\right)^{-1},~\label{geoeq2}
\end{eqnarray}
where the impact parameter is defined as $b=\sqrt{\frac{r_0^2}{g(r_0)}}$ with $r_0$ the radius of the circular orbit.

After introducing a new variable $u(\phi)=\frac{1}{r}$, the above geodesic equation can be rewritten as
\begin{eqnarray}
\left(\frac{du}{d\phi}\right)^2=\frac{1}{b^2}-u^2+2 M u^3\left[1-\tanh\left(\frac{Q_m^2 u}{2M}\right)\right],~\label{geo}
\end{eqnarray}
which can be solved by iterative method as follows:
\begin{eqnarray}
u(\phi)&=&\frac{\sin \phi }{b}+\frac{M \left(\cos ^2\phi +1\right)}{b^2}-\frac{M^2 \cos \phi}{8 b^3}  \Big(30 \phi \nonumber\\
       &+&3 \sin (2 \phi )-20 \tan \phi \Big)-\frac{Q_m^2\cos\phi}{2b^3}\bigg(-\frac{3 \phi }{2} \nonumber\\
       &+&\frac{1}{4} \sin (2 \phi )+\tan \phi \bigg)+\mathcal{O}(M^3,Q_m^3)~\label{uphi}.
\end{eqnarray}

Besides, the bending angle of light can be expressed as
\begin{eqnarray}
\hat{\alpha}(r_0)=\Delta\phi(r_0)-\pi,
\end{eqnarray}
where $\Delta\phi(r_0)$ can be obtained from Eq.~(\ref{geoeq2}) as
\begin{eqnarray}
\Delta\phi(r_0)&=&2\int_{r_0}^{\infty}\bigg(\frac{r^4}{b^2}-r^2 \nonumber\\
&+&2Mr\left[1-\tanh\left(\frac{Q_m^2}{2Mr}\right)\right]\bigg)^{-\frac{1}{2}}dr.
~\label{dphi2}
\end{eqnarray}

Defining the new dimensionless spacetime coordinates $x=\frac{r}{2M}$ and $T=\frac{t}{2M}$ and the dimensionless magnetic charge $q_m=\frac{Q_m}{2M}$, Eq.~(\ref{dphi2}) can be reexpressed as
\begin{eqnarray}
\Delta\phi(x_0)\!\!&=&\!\!2 \int_{x_0}^{\infty }\!\! x_0^{\frac{3}{2}} \Big(x^4(-1+x_0)+x x_0^3(1-x) \nonumber\\
\!\!&-&\!\!x x_0^3\tanh\left(q_m^2/x\right)\!+\!x^4\tanh\left(q_m^2/x_0\right)\Big)^{-\frac{1}{2}}\!\! dx, \nonumber\\
\end{eqnarray}
and the impact parameter is given by
\begin{eqnarray}
\frac{b}{2M}=\frac{x_0}{\sqrt{g(x_0)}}.~\label{im2}
\end{eqnarray}

After defining a new variable $z=\frac{x_0}{x}$, the above integral becomes
\begin{eqnarray}
\Delta\phi(x_0)\!\!&=&\!\!2 \int_0^1 \sqrt{x_0}\Big(-1+(1-z^2)x_0+z^3 \nonumber\\
\!\!&+&\!\!\tanh\left(q_m^2/x_0\right)\!-\!z^3\tanh\left(q_m^2 z/x_0\right)\!\Big)^{-\frac{1}{2}}\! dz.
\end{eqnarray}
Then the weak deflection angle can be integrated out term by term as follows:
\begin{widetext}
\begin{eqnarray}
\alpha(x_0)&=&\Delta\phi(x_0)-\pi=\frac{2}{x_0}+\left(\frac{\pi}{4} \left(\frac{15}{4}-3 q_m^2\right)-1\right)\frac{1}{x_0^2}+\left(\frac{3}{16} \pi  \left(4 q_m^2-5\right)-7 q_m^2+\frac{61}{12}\right)\frac{1}{x_0^3}+\bigg(\frac{5}{8} \left(20 q_m^2-13\right) \nonumber\\
&+&\frac{\pi  \left(320 q_m^6+912 q_m^4-6600 q_m^2+3465\right)}{1024}\bigg)\frac{1}{x_0^4}+\bigg(\frac{1}{120} (472-75 \pi )q_m^6+\frac{1}{32} (632-105 \pi )q_m^4 \nonumber\\
&+&\frac{7}{64} (135 \pi -536)q_m^2+\frac{7783}{320}-\frac{3465 \pi }{512}\bigg)\frac{1}{x_0^5}+\bigg(\frac{7 \pi  q_m^{10}}{48}+\frac{49 \pi  q_m^8}{64}-\frac{5}{96} (208+63 \pi )q_m^6-\frac{35}{1024} (2176+903 \pi )q_m^4 \nonumber\\
&+&\frac{9 }{4096}(70912+25305 \pi )q_m^2-\frac{21397}{384}-\frac{310695 \pi }{16384}\bigg)\frac{1}{x_0^6}+\mathcal{O}\left(\frac{1}{x_0^7}\right).~\label{alph2}
\end{eqnarray}
\end{widetext}

Besides, the relation between $b$ and $x_0$ can be obtained from Eq.~(\ref{im2}) as
\begin{eqnarray}
\frac{1}{x_0}&=&\frac{2M}{b}+\frac{1}{2}\left(\frac{2M}{b}\right)^2-\frac{1}{8}(4q_m^2-5)\left(\frac{2M}{b}\right)^3 \nonumber\\
&-&\left(\frac{3q_m^2}{2}-1\right)\left(\frac{2M}{b}\right)^4+\bigg(\frac{q_m^6}{6}+\frac{7q_m^4}{8}-\frac{63q_m^2}{16} \nonumber\\
&+&\frac{231}{128}\bigg)\left(\frac{2M}{b}\right)^5+\bigg(\frac{2q_m^6}{3}+5q_m^4-10q_m^2 \nonumber\\
&+&\frac{7}{2}\bigg)\left(\frac{2M}{b}\right)^6+\mathcal{O}\left(\left(\frac{2M}{b}\right)^7\right).~\label{x0b2}
\end{eqnarray}
Inserting Eq.~(\ref{x0b2}) into Eq.(\ref{alph2}), the weak deflection angle is found to be
\begin{eqnarray}
\hat{\alpha}\simeq\frac{4 M}{b}-\frac{3 \pi  Q_m^2}{4 b^2}-\frac{16MQ_m^2}{b^3}+\mathcal{O}(M^2,Q_m^3),
\end{eqnarray}
which is in agreement with the result calculated by using the Gauss-Bonnet theorem.

\subsection{Weak deflection angle in the presence of plasma}

In this subsection, we investigate the effects of a cold non-magnetized plasma on the deflection angle of the black hole in Einstein-Bronnikov theory. The refractive index for this black hole is given by \cite{Perlick2015},
\begin{eqnarray}
n(r)=\sqrt{1-\delta^2 g(r)}.
\end{eqnarray}
The corresponding optical metric is
\begin{eqnarray}
d\sigma^2&=&\gamma_{ij}dx^i dx^j=-\frac{n^2}{g_{00}}g_{ij}dx^i dx^j \nonumber\\
&=&n^2\left(\frac{1}{g^2}dr^2+\frac{r^2}{g}d\phi^2\right).~\label{oppmetric2}
\end{eqnarray}
Then the Gaussian curvature is calculated as
\begin{widetext}
\begin{eqnarray}
\tilde{\mathcal{K}}&=&\frac{1}{4r^4\delta^2}\bigg[-3r^2+4M\delta r-2\delta \text{sech}^2\left(\frac{Q_m^2}{2Mr}\right)\left(Q_m^2+Mr\sinh\left(\frac{Q_m^2}{Mr}\right)\right)\bigg] \nonumber\\
                   &+&\frac{6 (\delta -1) M r^3-2Q_m^2\delta\text{sech}^2\left(\frac{Q_m^2}{2Mr}\right)\left(Mr+Q_m^2\tanh\left(\frac{Q_m^2}{Mr}\right)\right)}{4M r^4\delta^2\left(r(1-\delta)+2M\delta\left(1-\tanh\left(\frac{Q_m^2}{2Mr}\right)\right)\right)} \nonumber\\
                   &+&\frac{1}{4M r^4\delta^2\left(r(1-\delta)+2M\delta\left(1-\tanh\left(\frac{Q_m^2}{2Mr}\right)\right)\right)^2}\bigg[Mr^4(-5+(8-3\delta)\delta)\nonumber\\
                   &+&2MQ_m^2r^2\delta(3\delta-2)\text{sech}^2\left(\frac{Q_m^2}{2Mr}\right)-Q_m^4\delta\text{sech}^4\left(\frac{Q_m^2}{2Mr}\right)\left(3M\delta+r\sinh\left(\frac{Q_m^2}{Mr}\right)\right)\bigg]\nonumber\\
                   &+&\frac{2r\left(r^2(\delta-1)-Q_m^2\delta\text{sech}^4\left(\frac{Q_m^2}{2Mr}\right)\right)^2}{4M r^4\delta^2\left(r(1-\delta)+2M\delta\left(1-\tanh\left(\frac{Q_m^2}{2Mr}\right)\right)\right)^3},
\end{eqnarray}
\end{widetext}
and the deflection angle can be obtained as
\begin{eqnarray}
\alpha&=&-\int^{\pi}_0\int^{\infty}_{\frac{1}{\sin\phi}}\tilde{\mathcal{K}}dS \nonumber\\
      &\simeq&\frac{2 M}{b}\left(1+\frac{1}{1-\delta }\right)-\frac{ \pi Q_m^2}{4 b^2}\left(1+\frac{2}{1-\delta}\right) \nonumber\\
      &+&\frac{2MQ_m^2}{b^3}\left(\frac{3\delta}{1-\delta}-\frac{8-10\delta}{(1-\delta)^2}\right)\nonumber\\
      &+&\mathcal{O}(M^2,Q_m^3),~\label{deflangp2}
\end{eqnarray}
It is obvious that Eq.~(\ref{deflangp2}) reduces to Eq.~(\ref{deflang2}) when $\delta\rightarrow 0$, and the deflection angle increases with the plasma parameter $\delta$, which suggests that the lower the photon frequency measured by a static observer at infinity is, the larger the deflection angle of it will be for a fixed electron plasma frequency.

\section{Conclusion}~\label{con}

As two well-known nonlinear electrodynamic (NLED) theories, Euler-Heisenberg NLED model and Bronnikov NLED model are extensively studied in the literatures.
In this paper, we considered the spherically symmetric and static black hole solutions based on these NLED models and calculated the weak deflection angle of light by these two black holes with the help of the Gauss-Bonnet theorem.
To be specific, in the Einstein-Euler-Heisenberg black hole, we investigated the effects of the one-loop corrections to quantum electrodynamics on the deflection angle of light and found that the weak deflection angle increases with the one-loop corrections. In the Einstein-Bronnikov black hole, we calculated the weak deflection angle by this regular magnetically charged black hole and found that the deflection angle by this black hole is smaller than the singular one. Besides, the weak deflection angles of both black holes were also calculated via the geodesic method, which was confirmed in agreement with the method by using the Gauss-Bonnet theorem at least at low order. What's more, the effects of a cold non-magnetized plasma on the weak deflection angle also were discussed and it was found that the deflection angle increases with the plasma parameter for both black holes, which indicates that the lower the photon frequency measured by a static observer at infinity is, the larger the deflection angle of it will be for a fixed electron plasma frequency.

\acknowledgments{
This work was supported by Scientific Research Program Funded by Shaanxi Provincial Education Department (No. 20JK0553), and the National Natural Science Foundation of China (Grants No. 11875151 and No. 11522541).
}

\end{document}